\documentclass[aps,prd,eqsecnum,preprint,tightenlines,nofootinbib,showpacs]{revtex4-1}
\usepackage{graphicx,amsmath,latexsym}

\newcommand{\bea}{\begin{eqnarray}}
\newcommand{\eea}{\end{eqnarray}}
\newcommand{\be}{\begin{equation}}
\newcommand{\ee}{\end{equation}}
\newcommand{\ub}[1]{\underline{#1}}
\newcommand{\ob}[1]{\overline{#1}}
\newcommand{\Pminus}{{\cal P}^-}

\begin{document}

\title{A nonperturbative light-front coupled-cluster method\footnote{Presented 
at QCD@Work2012, the International Workshop on QCD Theory and Experiment,
June 18-21, 2012, Lecce, Italy.}}

\author{J. R. Hiller}
\affiliation{Department of Physics \\
University of Minnesota-Duluth \\
Duluth, Minnesota 55812}

\date{\today}

\begin{abstract}
The nonperturbative Hamiltonian eigenvalue problem for bound states
of a quantum field theory is formulated in terms of Dirac's light-front
coordinates and then approximated by the exponential-operator technique 
of the many-body coupled-cluster method.  This approximation eliminates
any need for the usual approximation of Fock-space truncation.  
Instead, the exponentiated operator is truncated, and the terms retained
are determined by a set of nonlinear integral equations.  These
equations are solved simultaneously with an effective eigenvalue
problem in the valence sector, where the number of constituents
is small.  Matrix elements can be calculated, with extensions of
techniques from standard coupled-cluster theory, to obtain form
factors and other observables.
\end{abstract}


\maketitle


\section{Introduction}

\begin{figure}[b]
\includegraphics[width=11cm]{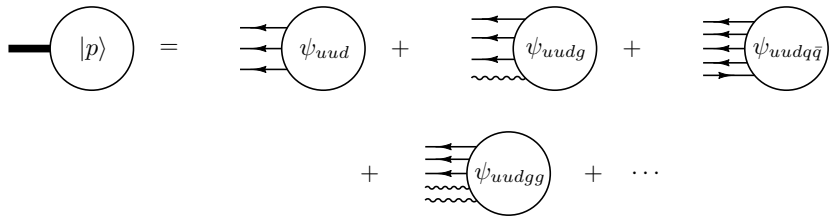}
\caption{\label{fig:proton} Fock-state expansion for
a proton in terms of quarks and gluons.}
\end{figure}

\begin{figure}
\includegraphics[width=14cm]{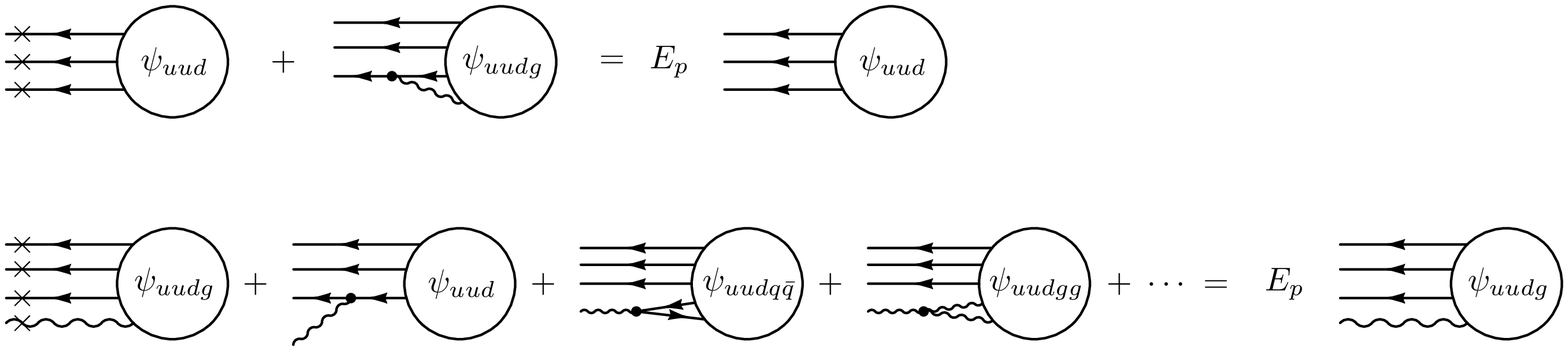}
\caption{\label{fig:coupledequations} Coupled equations for
the Fock-state wave functions of a proton.}
\end{figure}

We wish to construct a method to compute hadron
structure in terms of wave functions by solving
the light-front Hamiltonian eigenvalue problem
for quantum chromodynamics.  The use of light-front
quantization~\cite{Dirac} allows for a well-defined
Fock expansion~\cite{DLCQreviews}, as indicated
schematically in Fig.~\ref{fig:proton} for a proton.
The fundamental eigenvalue problem
\be
\left(\mbox{K.E.}+V_{\rm QCD}\right)|p\rangle=E_p|p\rangle,
\ee
with K.E. representing the kinetic energy,
$V_{\rm QCD}$ the interactions of QCD and
$E_p=\sqrt{m_p^2+p^2}$, is equivalent to a coupled
set of integral equations, depicted in Fig.~\ref{fig:coupledequations},
when projected onto Fock sectors.

In light-front quantization, the mass eigenvalue problem is
\be \label{eq:eigenvalueproblem}
\Pminus|\ub{P}\rangle=\frac{M^2+P_\perp^2}{P^+}|\ub{P}\rangle, \;\;\;\;
\underline{\cal P}|\ub{P}\rangle=\underline{P}|\ub{P}\rangle.
\ee
Here we define the light-front energy as $P^-=(E-P^z)$,
conjugate to the light-front time $x^+=t+z$; and the light-front
momentum $\ub{P}=(P^+\equiv E+P^z,\vec P_\perp=(P^x,P^y))$,
conjugate to the spatial coordinates $\ub{x}=(x^-\equiv t-z,x,y)$.
Because $p^+$ is positive for all constituents, there are
no spurious vacuum contributions to eigenstates;  particles
cannot be produced from the vacuum and still conserve $p^+$.
We also have a boost-invariant separation of internal
and external momenta.  Wave functions are then functions
of longitudinal momentum fractions $x_i\equiv p_i^+/P^+$
and relative transverse momenta 
$\vec k_{i\perp}\equiv \vec p_{i\perp}-x_i\vec P_\perp$
of the constituents.

To have a finite set of equations for the wave functions,
there must be a truncation of some sort.  The standard
truncation is one of Fock space, to limit the number
of constituents included in the calculation.  Unfortunately,
this introduces various complications, particularly
uncanceled divergences~\cite{SecDep}.  For example,
the Ward identity of gauge theories is destroyed
by the truncation because a truncation of the number
of photons in flight can eliminate vertex loops and
self-energy loops on one leg but not the other~\cite{OSUQED}.
We avoid these difficulties by making a different sort
of truncation, within the context of the light-front
coupled-cluster (LFCC) method~\cite{LFCClett,LC2011,QNP2012,LFCCqed}.

The LFCC method writes the eigenstate in the form
$|\psi\rangle=\sqrt{Z}e^T|\phi\rangle$, with $|\phi\rangle$
a valence state with the smallest number of constituents
consistent with the desired quantum numbers, $T$ an
operator that adds constituents, and $\sqrt{Z}$ a
normalization factor.  The operator $T$ conserves 
$J_z$, light-front momentum $\ub{P}$, charge and
any other quantum number that should be conserved.
We then define an effective Hamiltonian
$\ob{\Pminus}=e^{-T}\Pminus e^T$ and a projection
$P_v$ onto the valence sector, and the
original eigenvalue problem (\ref{eq:eigenvalueproblem})
is equivalent to
\be
P_v\ob{\Pminus}|\phi\rangle=\frac{M^2+P_\perp^2}{P^+}|\phi\rangle
\ee
and
\be
(1-P_v)\ob{\Pminus}|\phi\rangle=0.
\ee
The second equation is an auxiliary equation for the
operator $T$.

At this point, the formulation is still exact and still
reduces to an infinite set of equations for the infinite
set of terms possible for $T$.  The LFCC approximation
is a restriction of $T$ to one or a few terms and a
corresponding restriction of $1-P_v$ to generate only
enough auxiliary equations to solve for the terms
kept in $T$.  Many of the mathematical manipulations
needed to carry out this approach can be found in
the original many-body coupled-cluster 
method~\cite{CCorigin,CCreviews}.  There the valence
state is one of many particles, not a few, and the
$T$ operator introduces correlated excitations 
of the single-particle states without changing
particle number.  Despite the large difference
in the physics, the mathematics is much the same.

\section{Sample Application}

\begin{figure}
\includegraphics[height=1.5cm]{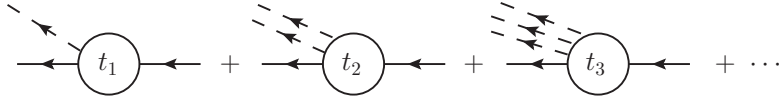}
\caption{\label{fig:T} The $T$ operator.  Each diagram
corresponds to a contribution to $T$ with emission of
one or more bosons, represented by dashed lines, from
the heavy fermion.  The circled vertex is labeled by
a function $t_n$ that determines the allocation of
momentum to each boson.}
\end{figure}

\begin{figure}
\includegraphics[height=4.5cm]{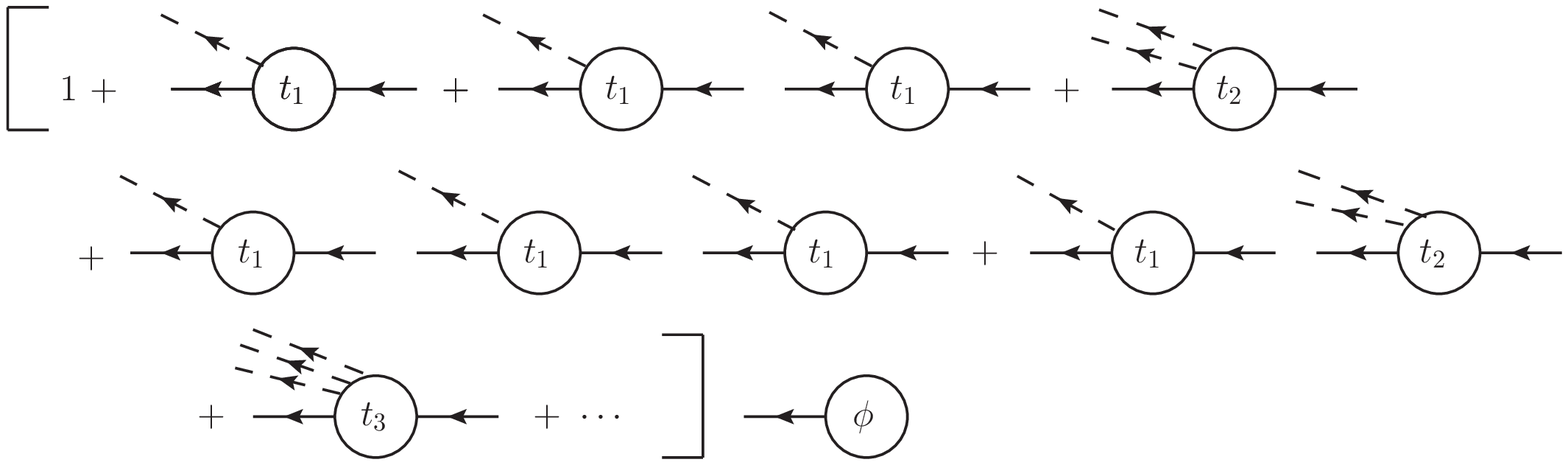}
\caption{\label{fig:eTphi} The action of $e^T$ on the
valence state $|\phi\rangle$.  Factorials and other
combinatoric factors are not shown.}
\end{figure}

\begin{figure}
\includegraphics[height=1.5cm]{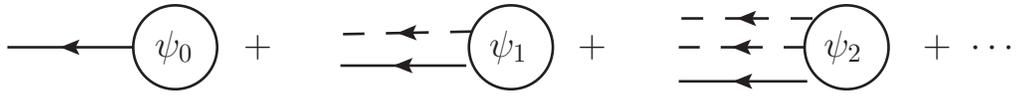}
\caption{\label{fig:psi} Fock-state expansion for
the eigenstate $|\psi\rangle$.  The circles are
labeled by the relevant wave function $\psi_n$.}
\end{figure}

To illustrate the use of the LFCC method
and to compare it with ordinary Fock-space
truncation, we consider
a light-front analog of the Greenberg--Schweber
model~\cite{GreenbergSchweber}.  In the model,
a heavy fermionic source emits and absorbs bosons
without changing its spin.  The full $T$ operator
would then be an expansion in multiple emissions
of bosons, as shown in Fig.~\ref{fig:T}.
The action of the exponentiation of $T$ is given in
Fig.~\ref{fig:eTphi}.
This is to be compared with a Fock-state expansion
of the eigenstate, as presented in Fig.~\ref{fig:psi},
where the wave function $\psi_0$ is equivalent to
the LFCC valence-state amplitude $\phi$.

\begin{figure}
\begin{tabular}{c}
\includegraphics[height=1.1cm]{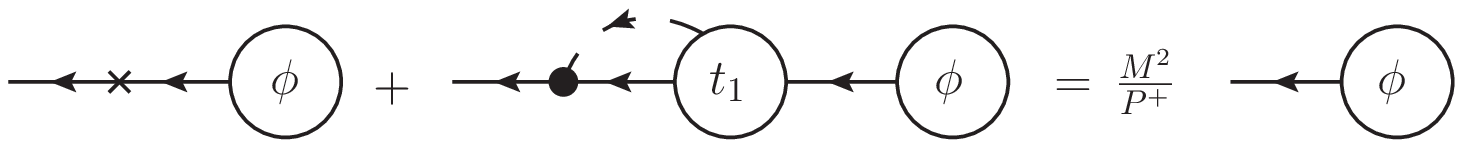} \\
\includegraphics[height=3.3cm]{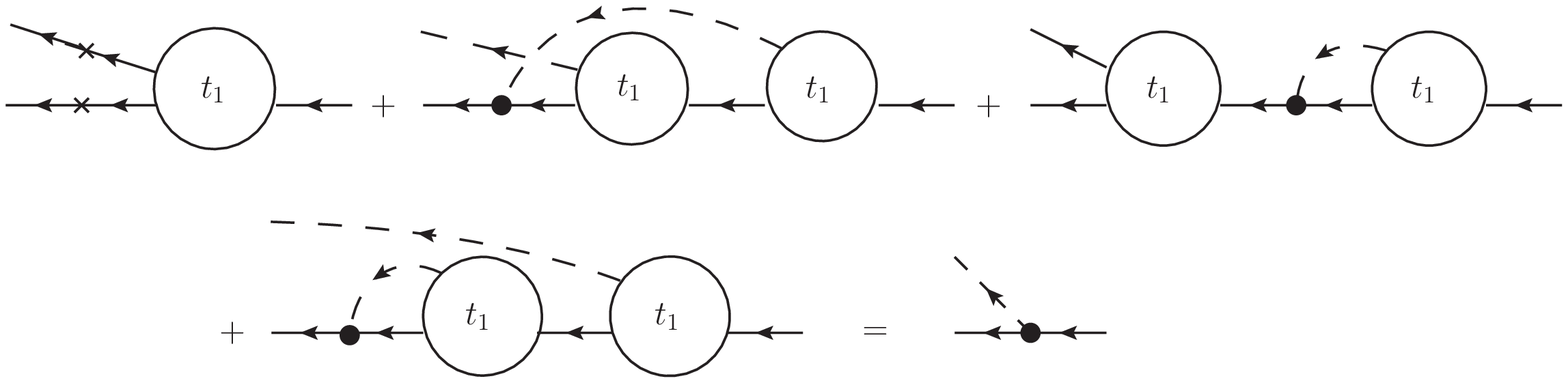}
\end{tabular}
\caption{\label{fig:LFCCeqn} Valence-sector eigenvalue
problem and auxiliary equation in the LFCC method.
The crosses represent kinetic energy contributions.
The inhomogeneous term on the right of the auxiliary 
equation is the bare emission vertex from the original
model Hamiltonian.}
\end{figure}

\begin{figure}
\includegraphics[height=3cm]{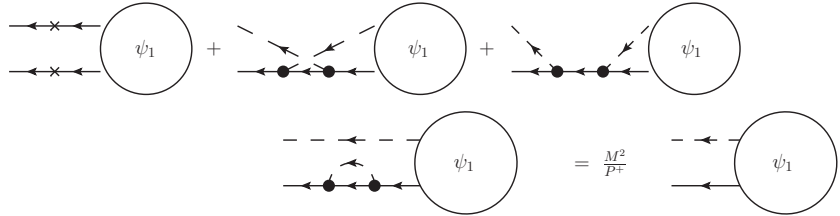}
\caption{\label{fig:psieqn} Equation for the one-boson/one-fermion
wave function $\psi_1$ obtained in a Fock-space truncation to
no more than two bosons.}
\end{figure}

A truncation of the $T$ operator to only the first
term, single-boson emission, leaves only the $t_1$
terms of Fig.~\ref{fig:eTphi}.  The valence
eigenvalue problem for $\phi$ and the auxiliary equation
for $t_1$ are then the coupled system of nonlinear
equations depicted in Fig.~\ref{fig:LFCCeqn}.
These are to be compared with the equation for
the Fock wave function $\psi_1$ in Fig.~\ref{fig:psieqn}
obtained by truncating the Fock-state expansion at two bosons
and then integrating out the zero-boson and two-boson
wave functions.
A key distinction between the two approaches is
in the self-energy loop for the fermion; in the LFCC
approach, it is the same everywhere it appears in Fig.~\ref{fig:LFCCeqn}.
For the Fock-space truncation approach, it is different
in each Fock sector, and the form that appears in Fig.~\ref{fig:psieqn}
is dependent on the momentum of the spectator boson.

In~\cite{LFCClett} we show that the simplest truncation of $T$ 
leads to an analytically solvable auxiliary equation
and provides for this simple model the exact answer!  The
solution can be used to compute matrix elements, in particular
the Dirac form factor for the fermion.~\footnote{The limitation
of the model to a heavy fermion that cannot flip its spin
eliminates consideration of the Pauli form factor; this is
not a limitation of the LFCC method.}

\section{Summary}

The LFCC method provides a nonperturbative approach to
the solution of Hamiltonian eigenvalue problems in quantum
field theories.  It assumes that the Hamiltonian is
suitably regulated, so that the integrals involved are
rendered finite.  The advantages of the method are that
it does not require a Fock-space truncation and that it
does not generate sector-dependent or spectator-dependent
self-energies.  Also, as an approximation, it is systematically
improvable through the addition of terms to the $T$ operator,
each constructed to add more particles to the valence state.

An application to a simple model has been discussed here
and more fully in \cite{LFCClett}.  An application to 
the dressed-electron state in QED is investigated in \cite{LFCCqed},
and additional work there is anticipated~\cite{LC2011,QNP2012}.
The method should also be applicable to QCD, with light-front
holographic QCD~\cite{hQCD} as a natural starting point;
there light-front holography may provide a useful approximation
to the LFCC valence eigenvalue problem for QCD.


\acknowledgments
This work was done in collaboration with S.S. Chabysheva
and supported in part by the US Department of Energy.


\begin{thebibliography}{}

\bibitem{Dirac} P.A.M. Dirac, 
\emph{Rev.\ Mod.\ Phys.}\ \textbf{21}, 392--399 (1949).
   
\bibitem{DLCQreviews} For reviews of light-cone quantization, see
   M. Burkardt, \emph{Adv.\ Nucl.\ Phys.}\ \textbf{23}, 1--74 (2002);
   S.J. Brodsky, H.-C. Pauli, and S.S. Pinsky, 
   \emph{Phys.\ Rep.}\ \textbf{301}, 299--486 (1998).
   
\bibitem{SecDep} S.S. Chabysheva and J.R. Hiller,
\emph{Ann.\ Phys.}\ \textbf{325}, 2435--2450 (2010).

\bibitem{OSUQED} D. Mustaki, S. Pinsky, J. Shigemitsu, and K. Wilson,
\emph{Phys.\ Rev.}\ D \textbf{43}, 3411--3427 (1991).
   
\bibitem{LFCClett} S.S. Chabysheva and J.R. Hiller,
\emph{Phys.\ Lett.}\ B \textbf{711}, 417--422 (2012).

\bibitem{LC2011} J.R. Hiller and S.S. Chabysheva,
\emph{Few Body Syst.}\ \textbf{52}, 315--321 (2012); 
S.S. Chabysheva and J.R. Hiller,
\emph{Few Body Syst.}\ \textbf{52}, 323--329 (2012).
  
\bibitem{QNP2012} J.R. Hiller,
\emph{PoS}(QNP2012), 113:1--6 (2012);
S. Chabysheva,
\emph{PoS}(QNP2012), 123:1--6 (2012).

\bibitem{LFCCqed} S.S. Chabysheva and J.R. Hiller,
``An application of the light-front coupled-cluster method
to the nonperturbative solution of QED,''
arXiv:1203.0250 [hep-ph].

\bibitem{CCorigin} F. Coester, 
\emph{Nucl.\ Phys.}\ \textbf{7}, 421--424 (1958);
F. Coester and H. K\"ummel, 
\emph{Nucl.\ Phys.}\ \textbf{17}, 477--485 (1960).

\bibitem{CCreviews} For reviews of the many-body coupled-cluster
method, see R.J. Bartlett and M. Musial, 
\emph{Rev.\ Mod.\ Phys.}\ \textbf{79}, 291--352 (2007);
T.D. Crawford, and H.F. Schaefer,
\emph{Rev.\ Comp.\ Chem.}\ \textbf{14}, 33--136 (2000);
R. Bishop, A.S. Kendall, L.Y. Wong, and Y. Xian,
\emph{Phys.\ Rev.}\ D \textbf{48}, 887--901 (1993);
H. K\"ummel, K.H. L\"uhrmann, and J.G. Zabolitzky,
\emph{Phys.\ Rep.}\ \textbf{36}, 1--63 (1978).

\bibitem{GreenbergSchweber} S.J. Brodsky, J.R. Hiller, and G. McCartor, 
\emph{Phys.\ Rev.}\ D \textbf{58}, 025005:1--16 (1998);
O. Greenberg, and S.S. Schweber,
\emph{Nuovo Cimento} \textbf{8}, 378--406 (1958).

\bibitem{hQCD} G.F.~de Teramond, and S.J.~Brodsky,
\emph{Phys.\ Rev.\ Lett.}\ \textbf{102}, 081601:1--4 (2009).

\end{thebibliography}
\end{document}